\documentclass[11pt,preprint]{aastex}

\usepackage{graphicx}
\usepackage{epsfig}

\shorttitle{TESTING THE NO-HAIR THEOREM: I. SPACETIME PROPERTIES}
\shortauthors{JOHANNSEN \& PSALTIS}

\begin{document}

\title{TESTING THE NO-HAIR THEOREM WITH OBSERVATIONS IN THE ELECTROMAGNETIC SPECTRUM:\\ I. PROPERTIES OF A QUASI-KERR SPACETIME}

\author{Tim Johannsen}
\affil{Physics Department,
University of Arizona,
1118 E. 4th Street,
Tucson, AZ 85721, USA}
\email{timj@physics.arizona.edu}

\author{Dimitrios Psaltis}
\affil{Astronomy Department,
University of Arizona,
933 North Cherry Avenue,
Tucson, AZ 85721, USA}
\email{dpsaltis@email.arizona.edu}

\begin{abstract}

According to the no-hair theorem, an astrophysical black hole is
uniquely described by only two quantities, the mass and the spin. In
this series of papers, we investigate a framework for testing the
no-hair theorem with observations of black holes in the
electromagnetic spectrum. We formulate our approach in terms of a
parametric spacetime which contains a quadrupole moment that is
independent of both mass and spin. If the no-hair theorem is correct,
then any deviation of the black-hole quadrupole moment from its Kerr
value has to be zero. We analyze in detail the properties of this
quasi-Kerr spacetime that are critical to interpreting observations of
black holes and demonstrate their dependence on the spin and
quadrupole moment. In particular, we show that the location of the
innermost stable circular orbit and the gravitational lensing
experienced by photons are affected significantly at even modest
deviations of the quadrupole moment from the value predicted by the
no-hair theorem. We argue that observations of black-hole images, of
relativistically broadened iron lines, as well as of thermal X-ray
spectra from accreting black holes will lead in the near future to an
experimental test of the no-hair theorem.

\end{abstract}


\section{INTRODUCTION}

The no-hair theorem establishes the remarkable property of general-relativistic black holes that their spacetimes and hence all of their characteristics are uniquely determined by precisely three parameters: their mass, spin, and charge (Israel 1967, 1968; Carter 1971, 1973; Hawking 1972; Robinson 1975; Mazur 1982). The spacetimes of such black holes are described by the Kerr-Newman metric (Newman et al. 1965), which reduces to the Kerr metric (Kerr 1963) in the case of uncharged black holes. This theorem relies on the cosmic censorship conjecture (Penrose 1969) as well as on the physically reasonable assumption that the exterior metric is free of closed timelike loops. If these requirements are met, then all astrophysical black holes should be described by the Kerr metric.

Black holes are commonly believed to be the final states of the evolution of sufficiently massive stars at the end of their lifecycle. The gravitational collapse of such stars leads to the formation of a black hole (Oppenheimer \& Snyder 1939; Penrose 1965), and any residual signature of the progenitor other than its mass and spin are radiated away by gravitational radiation (Price 1972a, 1972b). This scenario provides an astrophysical mechanism with which Kerr black holes can be generated. Almost all nearby galaxies harbor dark objects of high mass and compactness at their center (Kormendy \& Richstone 1995) including our own galaxy (Ghez et al. 2008; Gillessen et al. 2009) providing strong evidence that black holes are realized in nature. In addition, the measurement of the orbital parameters of many galactic binaries supports the claim that they contain stellar-mass black holes (e.g., McClintock \& Remillard 2006).

Despite the large amount of circumstantial evidence, there has been no direct proof, so far, for the existence of an actual event horizon. An event horizon, one of the most striking predictions of general relativity, is a virtual boundary that causally disconnects the interior of a black hole from the exterior universe. The presence of an event horizon in black-hole candidates has only been inferred indirectly, either from the properties of advection dominated accretion flows or from the lack of observations of Type I X-ray bursts (Narayan, Garcia, \& McClintock 1997, 2001; Narayan \& Heyl 2002; McClinock, Narayan, \& Rybicki 2004; Broderick et al. 2009; see discussion in Psaltis 2006).

The alternative hypothesis that these dark compact objects are not described by the Kerr metric but perhaps by a solution of the Einstein equations with a naked singularity (e.g., Manko \& Novikov 1992) and, therefore, violate the no-hair theorem, is still possible within general relativity. Alternatively, these dark objects might be stable stellar configurations consisting of exotic fields (e.g., boson stars, Friedberg, Lee, \& Pang 1987; gravastars, Mazur \& Mottola 2001; black stars, Barcel\'{o} et al. 2008). Finally, the fundamental theory of gravity may be different from general relativity in the strong-field regime, and the vacuum black-hole solution might not be described by the Kerr metric at all (e.g., Yunes \& Pretorius 2009; Sopuerta \& Yunes 2009; c.f. Psaltis et al. 2008). As a result, testing the no-hair theorem allows us not only to verify the identification of dark compact objects in the universe with Kerr black holes but to test the strong-field predictions of general relativity, as well.

The exterior spacetime of a black-hole candidate is usually best described in terms of its multipole moments. The mass and the spin of a black hole can be identified as the first two such moments. As a consequence of the no-hair theorem, mass and spin already specify all higher multipole moments of a black-hole spacetime completely. Therefore, measuring three independent multipole moments of the spacetime around a black hole suffices to test the no-hair theorem (Ryan 1995).

Recent advances in instrumentation have opened new horizon towards probing in detail the immediate vicinity of black holes. Very-long baseline interferometry (VLBI) has allowed us to observe the first images of the inner accretion flows of black holes and revealed evidence for sub-horizon scale structures (Doeleman et al. 2008). The International X-Ray Observatory (IXO) will measure the detailed profiles of relativistically broadened iron lines in AGN (see, e.g., Brenneman et al. 2009). Moreover, the Laser Interferometer Space Antenna (LISA) will probe supermassive black holes near their event horizons via extreme mass-ratio inspirals (EMRIs) (see, e.g., Hughes 2006).

On the theoretical front, several approaches have been developed to date in order to extract information about black-hole spacetimes with gravitational-wave observations. This can be done by analyzing several observables that are parametrized as functions of the multipole moments (Ryan 1995). It requires a suitable choice of the metric that either allows for a determination of multipole moments of order $l\geq2$ (Glampedakis \& Babak 2006; Gair, Li, \& Mandel 2008) or of metric perturbations (Collins \& Hughes 2004; Vigeland \& Hughes 2010). The principles of a general algorithm for stationary axisymmetric vacuum spacetimes that admit full integrability of the geodesic equations have been outlined by Brink (2008; 2009 and references therein).

In this series of papers, we investigate a framework for testing the no-hair theorem with observations in the electromagnetic spectrum. We formulate our tests in terms of a quasi-Kerr metric (Glampedakis \& Babak 2006) that incorporates an independent quadrupole moment but reduces smoothly to the familiar Kerr metric when the deviation at the quadrupole order is set to zero. The central object constitutes a quasi-Kerr black hole (similar to the term bumpy black hole coined by Collins \& Hughes 2004): it deviates from the Kerr metric in (at least) one multipole moment and allows for a test that distinguishes a Kerr black hole from a different type of object.

We compute observables that can be measured directly with either current or near-future instruments and allow for the extraction of at least three independent multipole moments. This approach allows for a two-fold test: If the central object is indeed a black hole, a deviation from the Kerr metric must be zero. If, however, the deviation is measured to be nonzero, then it either has to be a different type of object or general relativity itself breaks down in the strong-field regime very close to the black hole (c.f., Collins \& Hughes 2004; Hughes 2006).

In particular, our framework will enable us to perform a test of the no-hair theorem with the measurement of black-hole images, relativistically broadened iron lines emitted from accretion disks, and the measurement of the innermost stable circular orbit (ISCO) from the continuum disk spectra. All of these observables depend explicitly on the multipole moments of the spacetime. Under the assumption that the compact objects are Kerr black holes, these observables have been used to measure the spins of the black holes (e.g., Zhang, Cui, \& Chen 1997; Brenneman \& Reynolds 2006; Broderick et al. 2009). In the more general case, their properties and respective shapes likewise can be used to constrain the quadrupole moments. Similar approaches have been suggested based on timing observations of pulsar black hole binaries (Wex \& Kopeikin 1999) and on observations of stellar orbits in the vicinity of Sgr A* (Will 2008; Merritt et al. 2009).

This paper is structured as follows: In Section~2 we briefly summarize techniques to measure deviations from the Kerr metric and to extract multipole moments of a given black-hole spacetime. We develop our framework for a test of the no-hair theorem in Section~3 and analyze various relevant properties of the underlying quasi-Kerr metric in Section~4.

\section{MULTIPOLE MOMENTS AND THE NO-HAIR THEOREM}

In this section, we first discuss multipole expansions of curved spacetimes in general before we focus on the special nature of the Kerr metric. Then we briefly review four approaches to extracting observationally the spacetime parameters of a black hole.

\subsection{Multipole Expansions of Curved Spacetimes}

Symmetric systems in flat space are often best described by a set of multipole moments. In theories of gravity like general relativity, however, space is curved due to the presence of matter, and the resulting field equations are highly non-linear. It is, therefore, not immediately obvious that such a spacetime can actually be characterized by a set of multipole moments.

In Newtonian gravity, the potential $\Phi$ satisfies the Laplace equation
\begin{equation}
\nabla^2\Phi= \left\{ \begin{array}{ll}
              4\pi G\rho  & \mbox{(interior)}   \\
              0 & \mbox{(exterior)},\end{array} \right.
\end{equation}
where $G$ is Newton's constant, and $\rho$ is the mass density. Therefore, the potential $\Phi$ can always be expanded in spherical harmonics $Y_{lm}$ as
\begin{equation}
\Phi=-G\sum_{l=0}^{\infty}\frac{4\pi}{2l+1}\sum_{m=-l}^{l}\frac{M_{lm}Y_{lm}}{r^{l+1}}
\label{newtonexp}
\end{equation}
with mass multipole moments
\begin{equation}
M_{lm}=\int_0^r r'^{l+2}dr'\oint d\Omega' Y_{lm}^*(\Omega')\rho(r',\Omega').
\label{newtonmult}
\end{equation}
In these expressions, we have left the gravitational constant $G$ explicit. We will set this constant as well as the speed of light $c$ to unity for the remainder of this paper.

In the curved space of general relativity, however, the vacuum Einstein equations
\begin{equation}
R_{\mu\nu}-\frac{1}{2}g_{\mu\nu}R=0
\label{einsteinvac}
\end{equation}
have to be solved for the spacetime metric $g_{\mu\nu}$ with the corresponding Ricci tensor $R_{\mu\nu}$ and Ricci scalar $R$. The Einstein equations are nonlinear and, therefore, cannot always be solved in terms of an expansion over orthonormal polynomials. Nonetheless, it can be shown that a multipole expansion of curved spacetime does indeed exist (see Thorne 1980 for a detailed review).

For an asymptotically flat vacuum solution of the Einstein equations, (tensor) multipole moments can be defined based on a conformal compactification of 3-space, if the spacetime is also static (Geroch 1970) or, more generally, if it is stationary (Hansen 1974). In both cases, such a set of multipole moments characterizes the spacetime uniquely (Beig \& Simon 1980, 1981; see also Hauser \& Ernst 1981 and references therein) and obeys an appropriate convergence condition (B\"ackdahl \& Herberthson 2006). If the spacetime is also axisymmetric, the multipole moments are given by a bi-infinite series of scalars $M_l$ and $S_l$, which are interpreted as mass and current multipole moments, respectively. The mass multipole moments are analogous to the multipole moments in Newtonian gravity given by expression (\ref{newtonmult}) and are nonzero only for even $l$. The current multipole moments are nonzero only for odd $l$ and arise from the fact that, in general relativity, all forms of energy gravitate (Hansen 1974). Stationary axisymmetric vacuum solutions of the Einstein equations can likewise be generated from a given set of multipole moments (Sibgatullin 1991; Manko \& Sibgatullin 1993).

In general relativity, black-hole spacetimes are asymptotically flat vacuum solutions of the Einstein equations. These spacetimes must also be axisymmetric (Hawking 1972) and can therefore be described by a sequence of multipole moments $\{M_l,S_l\}$. As a consequence of the no-hair theorem, the Kerr spacetime is the unique black-hole solution within general relativity that contains an event horizon, and all multipole moments of order $l\geq2$ are determined only by the first two, i.e., by the mass $M=M_0$ and the spin $J=S_1$. This fact can be expressed mathematically with the relation (Geroch 1970; Hansen 1974)
\begin{equation}
M_{l}+{\rm i}S_{l}=M({\rm i}a)^{l},
\label{kerrmult}
\end{equation}
\noindent where $a=J/M$ is the dimensionless spin parameter.

In the astrophysical context, the fact that the no-hair theorem requires the multipole moments to be locked by expression (\ref{kerrmult}) allows for it to be tested quantitatively using observations of black holes. Since the first two multipole moments (i.e., the mass and spin) already specify the {\it entire} spacetime, any observable can ultimately depend only on those two moments. Therefore, a promising strategy for testing the no-hair theorem is to measure (at least) three multipole moments of the spacetime of a black hole (Ryan 1995).

\subsection{Four Approaches to Measuring the Multipole Moments of Curved Spacetimes}

To date, four approaches to testing the no-hair theorem have been suggested that are based on either a multipole expansion or a perturbation of the Kerr spacetime. In the following, we briefly review these methods in order to identify the one that is most suitable for our purposes.

The idea to ``map'' the spacetime of an astrophysical black hole and, therefore, to directly measure its multipole moments was first developed by Ryan (1995). He used an expansion in Geroch-Hansen multipoles in order to design a detection mechanism that could, in principle, probe the spacetime of a black hole using extreme mass-ratio inspiral (EMRI) observations. He showed that the values of the multipoles are encoded in several observables that can be measured by future experiments. In particular, these values can be extracted from the evolution of the orbital phase of inspiraling objects (by, e.g., X-ray timing missions) or from the gravitational wave spectrum using LISA (Ryan 1995, 1997a, 1997b).

Ryan (1995) considered a spacetime that is stationary, axisymmetric, and asymptotically flat. The inspiraling object moves gradually along a family of geodesic orbits in the equatorial plane and radiates away its energy and angular momentum adiabatically. The metric takes the form (Papapetrou 1953)
\begin{equation}
ds^2=-F(dt-\omega d\phi)^2 + \frac{1}{F}[e^{2\gamma}(d\rho^2+dz^2)+\rho^2d\phi^2],
\label{papapetrou}
\end{equation}
\noindent using cylindrical Weyl coordinates ($t,~\rho,~\phi,~z$). The functions $F$, $\gamma$, and $\omega$ depend on the coordinates $\rho$ and $z$ and can be generated from the Ernst potential (Ernst 1968)
\begin{equation}
\mathcal{E}=F+i\Psi=\frac{\sqrt{\rho^2+z^2}-\xi}{\sqrt{\rho^2+z^2}+\xi}.
\end{equation}
\noindent The function $\xi$ can be expanded as (Fodor, Hoenselaers \& Perj\'es 1989)
\begin{equation}
\xi=\sum_{j,k=0}^{\infty}a_{jk}\frac{\rho^j z^k}{(\rho^2+z^2)^{j+k}},
\label{xitilda}
\end{equation}
\noindent which specifies $F(\rho,z)$. The coefficients $a_{jk}$ are directly related to the multipole moments of the spacetime given by the metric in expression (\ref{papapetrou}). The remaining functions $\omega(\rho,z)$ and $\gamma(\rho,z)$ are then obtained in the form of two integrals (Dietz 1984; Wald 1984, pp. 165-67). Barack \& Cutler (2004; 2007) used approximate waveforms to extend Ryan's analysis for generic orbits and included modulations caused by LISA satellites. Li \& Lovelace (2008) generalized Ryan's approach to include tidal coupling between the inspiraling and the central object. One might interject at this point (see also Collins \& Hughes 2004) that a $1/r$-expansion such as the one exhibited by equation (\ref{xitilda}) spoils the promise of an approach that is based on the multipole moments of the spacetime. The potential difficulty is that a large number of such moments will be required to describe the spacetime as one approaches the event horizon.

In an alternative approach, Collins \& Hughes (2004) constructed spacetimes of perturbed Schwarzschild black holes (bumpy black holes). These spacetimes have the usual symmetries of being stationary, axisymmetric, and asymptotically flat, and the black hole is effectively perturbed by the presence of additional fields exterior to the black-hole horizon. For these configurations Collins \& Hughes (2004) calculated the periapse precession of equatorial orbits and found a significant (factor $\sim10$) increase of the precession in the strong-field regime.

Their starting point was the Weyl metric (Weyl 1918)
\begin{equation}
ds^2=-e^{2\psi}dt^2+e^{2\gamma-2\psi}(d\rho^2+dz^2)+e^{-2\psi}\rho^2d\phi^2,
\end{equation}
\noindent which is identical to the metric (\ref{papapetrou}) with the substitutions $F=e^{2\psi}$ and $\omega=0$. The quantities $\psi$ and $\gamma$ are solutions to the vacuum Einstein equations (Collins \& Hughes 2004; see also Suen, Price \& Redmount 1988).

Then Collins \& Hughes (2004) introduced small perturbations as $\psi=\psi_0+\psi_1$ and $\gamma=\gamma_0+\gamma_1$, where $\psi_0$ and $\gamma_0$ are the corresponding expressions of the Schwarzschild metric, and solved the Einstein equations keeping only leading order terms in the perturbation. (Note that this technique does not make explicit use of the multipole moments.) This approach was recently generalized by Vigeland \& Hughes (2010) who considered similar perturbations in the metric of rotating black holes.

A third approach is based on the Manko-Novikov metric (Manko \& Novikov 1992). This metric is stationary, axisymmetric, and asymptotically flat and is also a solution of the vacuum Einstein equations. It is a parametric extension of the Kerr metric with a different set of multipole moments. In this metric, mass and current multipole moments are coupled. In addition to the usual first two moments, i.e., mass and spin, the mass multipole moments of order $l\geq2$ are free parameters. They in turn specify the current multipole moments. In general, this metric has no event horizon but contains a naked singularity at the center (Manko \& Novikov 1992). Gair et al. (2008) studied a subclass of the Manko-Novikov metric analyzing possible observational signatures of ergodic non-Kerr EMRIs. Apostolatos, Lukes-Gerakopoulos, \& Contopoulos (2009) showed that the appearance of Birkhoff chains in the neighborhood of resonant tori (due to the Poincar\'e-Birkhoff theorem, Poincar\'e 1912; Birkhoff 1913) would lead to such a modification of a gravitational-wave measurement.

Finally, in a fourth approach, Glampedakis \& Babak (2006) considered perturbations of the Kerr metric that still satisfy the vacuum Einstein equations by exploiting the properties of the Hartle-Thorne metric (Hartle 1967; Hartle \& Thorne 1968). This metric, in Hartle-Thorne coordinates, describes the spacetime outside any slowly rotating compact object within general relativity and was originally designed for neutron stars. It provides an expansion of a general spacetime up to the quadrupole order and has a quadrupole moment $Q$ that is independent of both mass and spin (Hartle \& Thorne 1968). Glampedakis \& Babak (2006) constructed a quasi-Kerr extension to the Kerr metric by choosing the (dimensionless) quadrupole moment to be $q_{\rm Kerr}-\epsilon$, where $\epsilon$ parametrizes a potential deviation from the Kerr metric and $q_{\rm Kerr}=-J^2/M^4=-a^2/M^2$. In this parametrization, the quadrupole moment is given by (Glampedakis \& Babak 2006)
\begin{equation}
Q=-M\left(a^2+\epsilon M^2\right).
\label{qradmoment}
\end{equation}
We verified explicitly that this metric is a solution of the vacuum Einstein equations up to the quadrupole order by showing that the Ricci tensor vanishes.

The quasi-Kerr metric is stationary, axisymmetric, and asymptotically flat and reduces to the Kerr metric in the limit $\epsilon\rightarrow0$. The particular choice of Boyer-Lindquist coordinates by Glampedakis \& Babak (2006) is convenient, because it preserves the special Petrov-type D character of the Kerr part. Similar approaches have also been analyzed in the context of neutron stars (Laarakkers \& Poisson 1999; Berti \& Stergioulas 2004), rigidly rotating stars (Bradley \& Fodor 2009), and naked singularities (Bini et al. 2009).

For equatorial orbits in the quasi-Kerr metric, Glampedakis \& Babak (2006) calculated the periastron precession and constructed `kludge' gravitational waveforms as a function of the parameter $\epsilon$. These waveforms can be significantly different from the expected Kerr signal even for small changes of the quadrupole moment. They identified, however, a confusion problem, because it is possible to match quasi-Kerr waveforms of one set of orbital parameters with the Kerr template of a different set of orbital parameters.

The fact that the quasi-Kerr metric is a valid solution of the Einstein equations even for non-Kerr values of the quadrupole moment allows us to design a self-consistent test of the no-hair theorem and of the nature of the compact object within general relativity. A disadvantage of the quasi-Kerr metric is the restriction that it cannot be used for a description of rapidly spinning black holes. The most promising approach will be, of course, to perform the tests of the no-hair theorem with all of the above parametrizations of spacetime in order to explore the robustness of the results. Nonetheless, the fact that the quasi-Kerr metric contains an independent quadrupole multipole moment makes it a natural and well-suited environment for a test of the no-hair theorem, which we adopt below.

\section{A FRAMEWORK FOR TESTING THE NO-HAIR THEOREM}

In this section, we formulate our framework for testing the no-hair theorem with observations of black holes in the electromagnetic spectrum. We use the quasi-Kerr metric (Glampedakis \& Babak 2006) as the underlying spacetime and develop a method from which we derive observables parametrized by an independent quadrupole moment that is potentially different from the value predicted by the no-hair theorem.

We name the central object described by such a metric a quasi-Kerr black hole (similar to the term bumpy black hole, c.f. Collins \& Hughes 2004). This is a black hole within general relativity only if the deviation of all its multipole moments from the respective values in the Kerr spacetime are zero. Similarly to the case of EMRI observations (Collins \& Hughes 2004; Hughes 2006), our approach may constitute a null-hypothesis test: If the central object is indeed a general-relativistic black hole, then it must have the multipole structure of the Kerr metric given by relation (\ref{kerrmult}). If a deviation of the multipole moments from this expression is detected, the object cannot be a general-relativistic black hole. In that case, the object is either a star or, if it is otherwise known to posess an event horizon, a black hole with a different set of multipole moments, which, therefore, violates the no-hair theorem. Contrary to gravitational-wave analyses, however, which rely explicitly on the Einstein equations, our framework requires only the metric and the Einstein Equivalence Principle. Thus, it can also be viewed as a self-consistent test of general relativity in the strong-field regime.

In the following, we specify the exact form of the quasi-Kerr spacetime and we estimate its range of applicability. This metric has three independent multipole moments (mass, spin, and quadrupole moment). The deviation of the quadrupole moment from the Kerr value is written in terms of the dimensionless parameter $\epsilon$, so that the full quadrupole moment is given by relation (\ref{qradmoment}).

In Boyer-Lindquist coordinates, the Kerr metric $g_{\rm ab}^{\rm K}$ takes the form (e.g., Bardeen, Press, \& Teukolsky 1972)
\[
ds^2=-\left(1-\frac{2Mr}{\Sigma}\right)~dt^2-\left(\frac{4Mar\sin^2\theta}{\Sigma}\right)~dtd\phi
\]
\begin{equation}
+\left(\frac{\Sigma}{\Delta}\right)~dr^2+\Sigma~d\theta^2+\left(r^2+a^2+\frac{2Ma^2r\sin^2\theta}{\Sigma}\right)\sin^2\theta~d\phi^2
\label{kerr}
\end{equation}
\noindent with
\[
\Delta\equiv r^2-2Mr+a^2,
\]
\begin{equation}
\Sigma\equiv r^2+a^2\cos^2~\theta.
\label{deltasigma}
\end{equation}

The quadrupolar correction is determined by choosing a quadrupole moment of the form (\ref{qradmoment}) in the Hartle-Thorne metric (Hartle \& Thorne 1968). Then, the quasi-Kerr metric $g_{\rm ab}^{\rm QK}$ in Boyer-Lindquist coordinates is given by (Glampedakis \& Babak 2006)
\begin{equation}
g_{\rm ab}^{\rm QK}=g_{\rm ab}^{\rm K}+\epsilon h_{\rm ab}.
\label{qKerr}
\end{equation}
\noindent
In contravariant form, $h^{\rm ab}$ is
\[
h^{\rm tt}=(1-2M/r)^{-1}\left[\left(1-3\cos^2\theta\right)\mathcal{F}_1(r)\right],
\]
\[
h^{\rm rr}=(1-2M/r)\left[\left(1-3\cos^2\theta\right)\mathcal{F}_1(r)\right],
\]
\[
h^{\rm \theta\theta}=-\frac{1}{r^2}\left[\left(1-3\cos^2\theta\right)\mathcal{F}_2(r)\right],
\]
\[
h^{\rm \phi\phi}=-\frac{1}{r^2\sin^2\theta}\left[\left(1-3\cos^2\theta\right)\mathcal{F}_2(r)\right],
\]
\begin{equation}
h^{\rm t\phi}=0.
\end{equation}
The functions $\mathcal{F}_{1,2}(r)$ are given in Appendix A of Glampedakis \& Babak (2006). Note that, although only the quadrupole moment of the spacetime has been altered, the unperturbed spacetime is formally correct up to the maximum value of the spin.

In the case of the regular Kerr metric (in Boyer-Lindquist coordinates), its Petrov-type D character ensures the existence of four constants of motion: the particle rest mass $\mu$, its energy $E$, the angular momentum about the z-axis $L_z$, and the Carter constant (Carter 1968). Equivalently, the Hamilton-Jacobi equation is separable in all four coordinates, i.e., time $t$, radius $r$, and angles $\theta$ and $\phi$, and the equations of motion are reduced to quadratures (see, e.g., Chandrasekhar 1983). This special property greatly simplifies the analysis of a variety of astrophysical applications in the Kerr geometry.

The quasi-Kerr metric in contrast admits full separability of the Hamilton-Jacobi equation only for equatorial orbits. For generic orbits, the Carter constant is lost. An approximate solution can be found in the neighborhood of the central object using the action-angle formalism of canonical perturbation theory (Glampedakis \& Babak 2006). Unfortunately, this is no longer possible for trajectories that escape to infinity, such as those of photons that reach a distant observer. Consequently, the equations of motion have to be solved numerically. Berti et al. (2005) pointed out quantitatively how a deviation from the Kerr quadrupole moment alters the Petrov type of a spacetime.

Before embarking into the study of the properties of the quasi-Kerr metric that are relevant for observational tests of the no-hair theorem, it is important to check that the multipole expansion of the metric is well-behaved at small radii. An approach based on a multipole expansion breaks down in the limit $r\rightarrow0$, and a higher number of multipole moments must be kept in order for the expansion to accurately describe the black-hole spacetime near the event horizon (Ryan 1995; Collins \& Hughes 2004). The spacetime at horizon scales, however, is most interesting for tests of the no-hair theorem. In the following, we estimate the range of validity of the quasi-Kerr metric. A formal bound on the multipole moments of stationary spacetimes can be found in B\"ackdahl \& Herberthson (2006), but this is beyond the scope of this paper.

The validity of the expansion of the metric considered above breaks down at a critical radius $r_{\rm c}$ that depends both on the spin $a$ and the parameter $\epsilon$. This cutoff occurs at the radius at which terms of order $\epsilon^2$ and $\epsilon a$ can no longer be neglected against the correction terms proportional to $\epsilon$. Unfortunately, these higher order terms cannot be calculated uniquely in the approach of Glampedakis and Babak (2006). In the Kerr metric, the expansion of the diagonal elements in the spin parameter $a$ has nonvanishing corrections only at order $a^4$. These terms are always smaller (in magnitude) than the terms at order $a^2$ for $r>2M$ and are negligible at radii $r\gtrsim2.5M$. Therefore, in order to estimate the value of the critical radius $r_{\rm c}$, we compare the quadrupolar correction terms of order $\epsilon$ in the quasi-Kerr metric given by expression (\ref{qKerr}) to the respective elements of the Kerr metric in expression (\ref{kerr}) up to the quadrupole order.

The metric element which is affected most by the correction that is linear in the parameter $\epsilon$ is $g_{\rm rr}^{\rm QK}$. In Figure~1, we plot the radius at which the absolute value of the quadrupolar correction equals (solid lines) the absolute value of the Kerr element $g_{\rm rr}^{\rm K}$ up to order $a^2$ as a function of the spin $a$. At this radius, the element $g_{\rm rr}^{\rm QK}$ changes sign if $\epsilon>0$. This unphysical property renders the metric non-Lorentzian and leads to a reflection of infalling particles.\footnote{We thank A. Broderick for pointing this out.} We estimate the cutoff $r_{\rm c}$ as the radius at which the absolute value of the quadrupolar correction equals $50\%$ of the absolute value of the Kerr element $g_{\rm rr}^{\rm K}$ up to order $a^2$ (Figure~1, dashed lines). Note that these validity constraints are necessary but not sufficient.

\begin{figure}[t*]
\label{validity}
\begin{center}
\psfig{figure=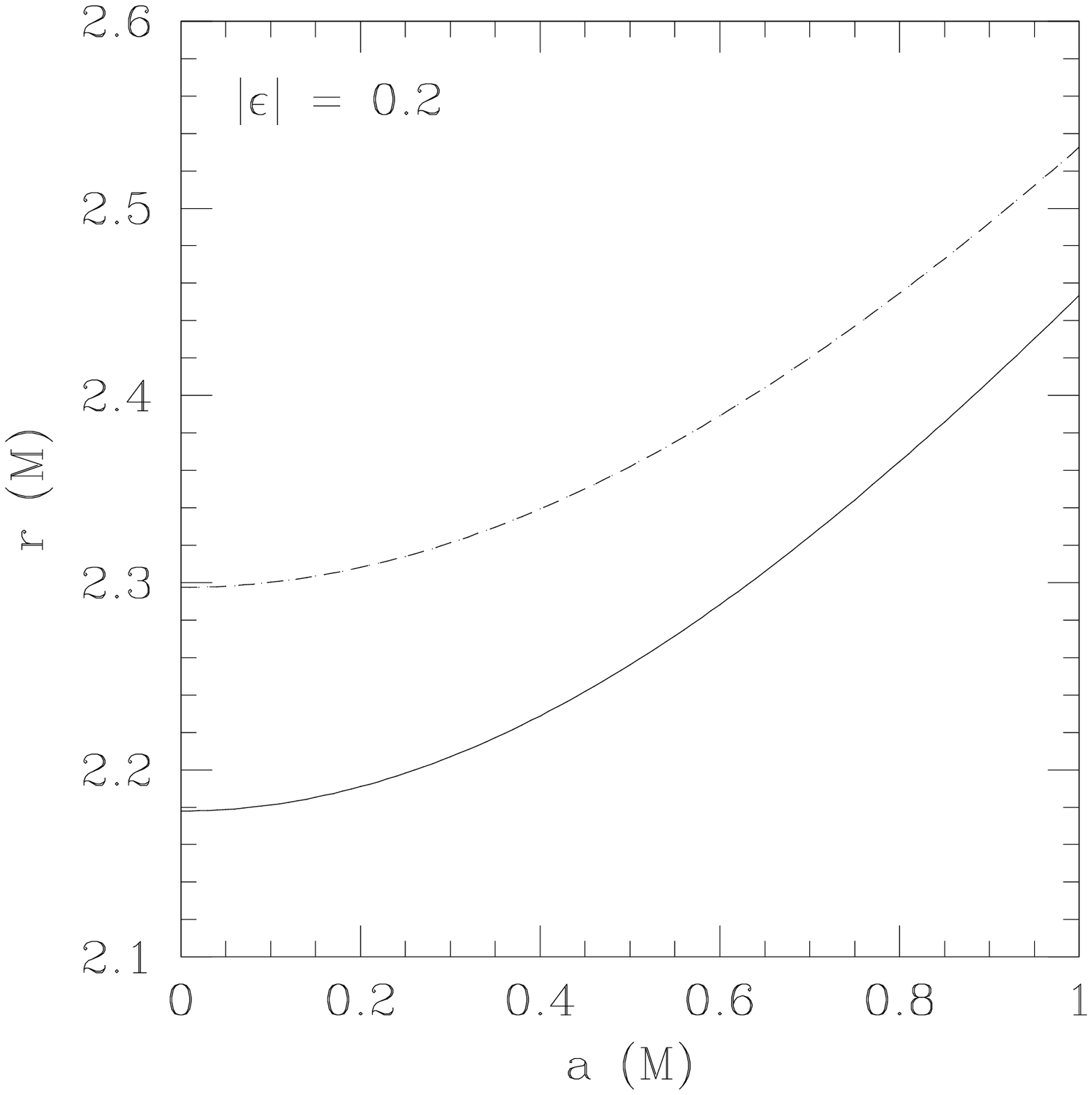,height=2in}
\psfig{figure=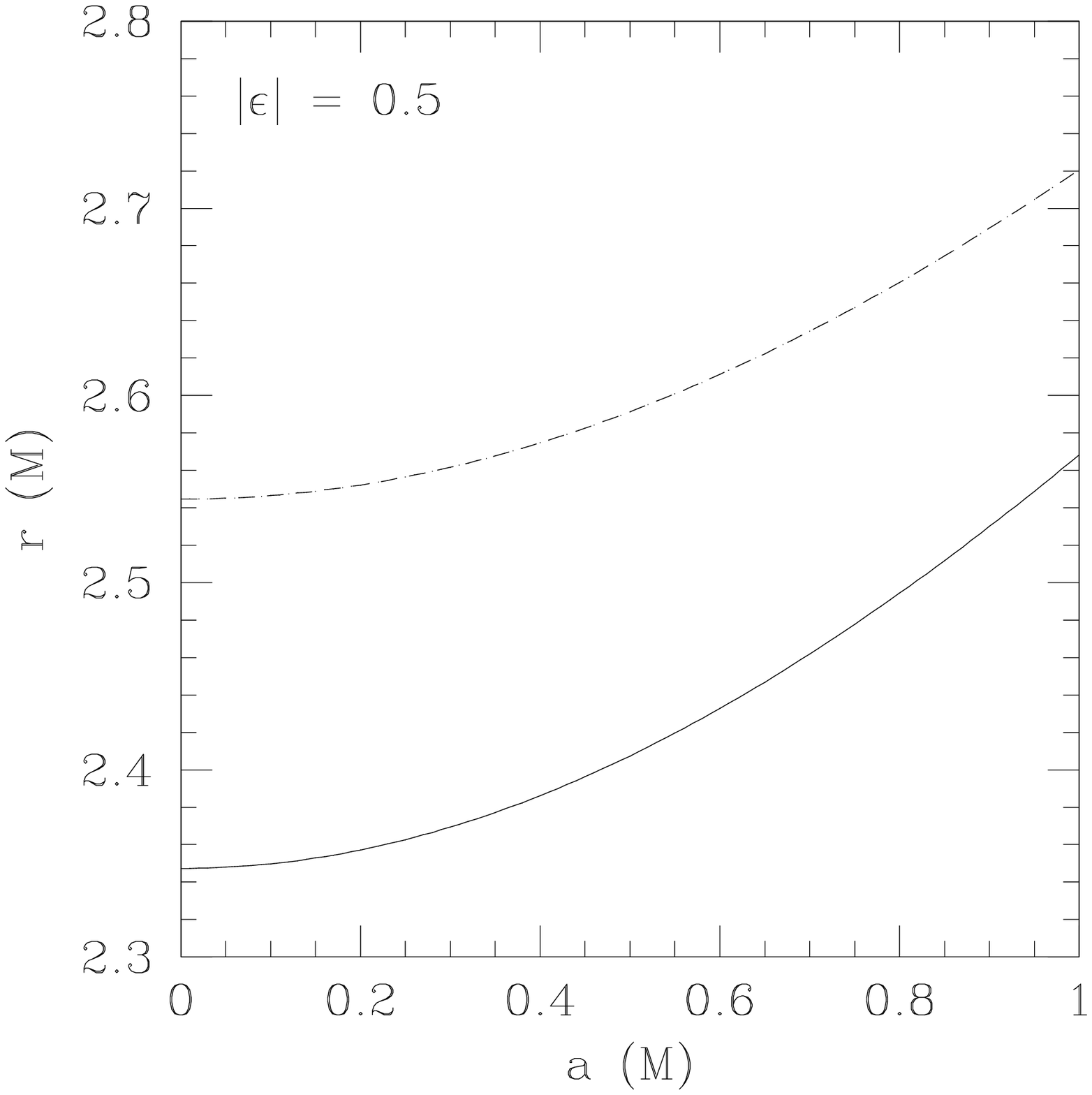,height=2in}
\psfig{figure=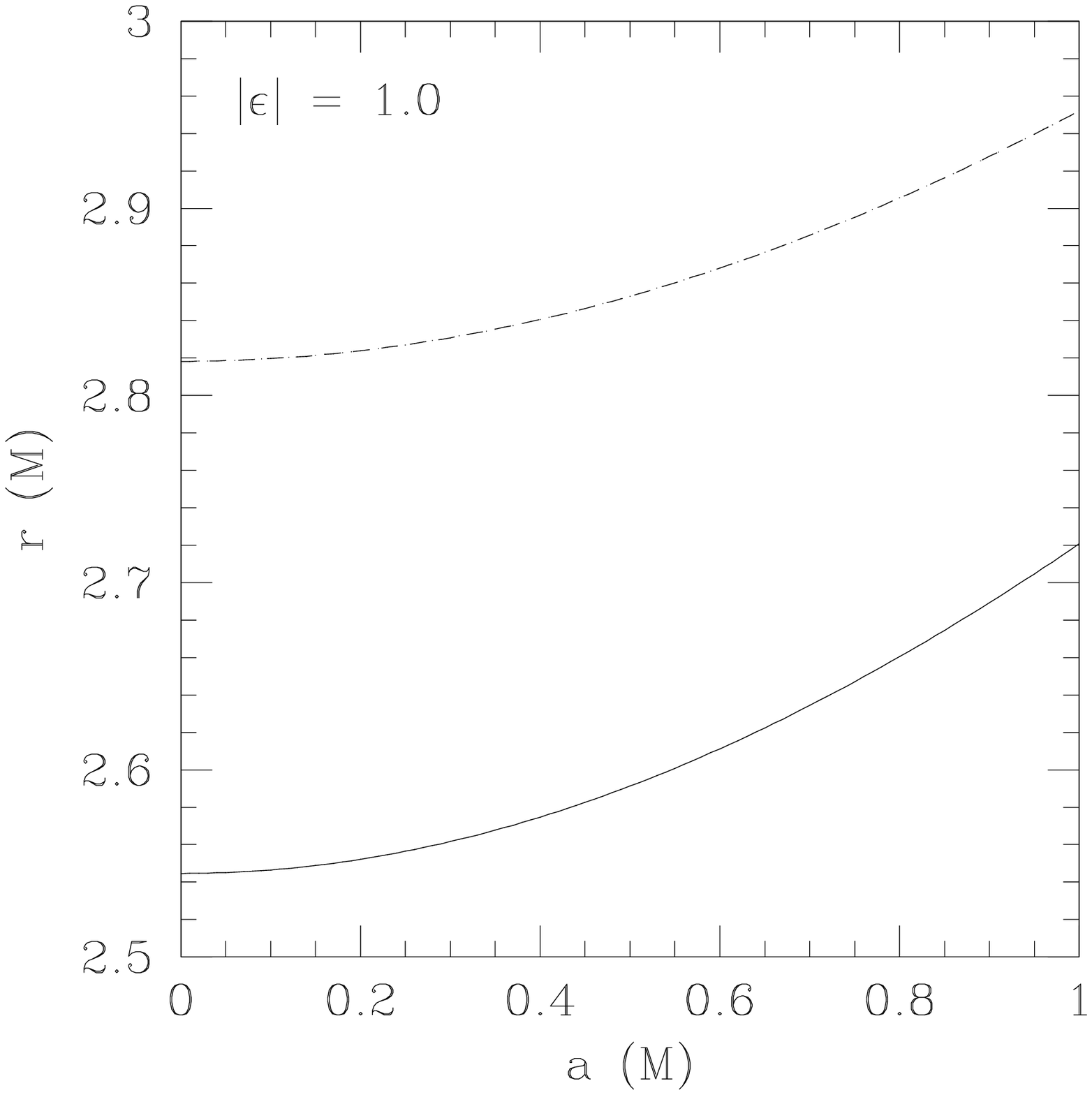,height=2in}
\end{center}
\caption{The radius $r$ at which the absolute value of the quadrupolar correction of the element $g_{\rm rr}^{\rm QK}$ is equal to (solid lines) $100\%$ and (dashed lines) $50\%$ of the absolute value of the element $g_{\rm rr}^{\rm K}$ expanded to order $a^2$. At the radii along the solid lines, the metric element $g_{\rm rr}^{\rm QK}$ changes sign if $\epsilon>0$ and the metric becomes non-Lorentzian.}
\end{figure}

The excluded domain $r<r_{\rm c}$ lies inside the circular photon orbit of the Kerr metric for $a\lesssim0.4$ and $\epsilon\lesssim0.5$ (c.f., Bardeen et al. 1972). Therefore, in the quasi-Kerr metric, we expect the fraction of photons from this region that reach an observer at infinity to be negligible. The only exception are photons from further out in the accretion disk that are strongly lensed by the black hole and approach the event horizon closely.

While higher order terms of the spin $a$ in the quasi-Kerr metric are negligible for values of the radius $r\gtrsim2.5M$, this may not be the case for terms of order $\epsilon a$ if the spin parameter $a$ is too large. We will, therefore, limit our discussion to values of the spin $a\leq0.4$. For black holes with this amount of spin and a value of the parameter $\epsilon\approx0.5$, the critical radius $r<r_{\rm c}$ is comparable to the radius of the circular photon orbit in the Kerr metric.

An additional complication arises from a singularity at $r_{\rm s}=2M$. This singularity is a relic of the expansion in the Hartle-Thorne metric. It vanishes in the Kerr metric, where all divergences cancel, hence illustrating the peculiar nature of that metric. Since the singularity is located at $r_{\rm s}<r_{\rm c}$, it plays only a very minor role in the following discussion. It does, however, effect the estimate of the critical radius $r_{\rm c}$ and therefore would make a different choice of the coordinates desirable.

Technically speaking, our estimate for the critical radius $r_{\rm c}$ is only valid for the elements of the quasi-Kerr metric $g_{\rm ab}^{\rm QK}$, but not for its derivatives. In the case of large quadrupolar corrections, $|\epsilon|\gtrsim a^2$, the corresponding Christoffel symbols depend significantly on the value of the parameter $\epsilon$ at radii larger than the critical radius, and the expansion to linear order in the parameter $\epsilon$ is no longer perturbative.

Our approach hereafter is to numerically integrate the geodesic equation in the quasi-Kerr spacetime for photons that are emitted from an accretion flow around the black hole and reach a distant observer. The geodesic equation takes the form
\begin{equation}
\frac{d^2x^{\rm \alpha}}{d\lambda^2}=-\Gamma^{\rm \alpha}_{\rm \beta\gamma}\frac{dx^{\rm \beta}}{d\lambda}\frac{dx^{\rm \gamma}}{d\lambda},
\label{geodesiceq}
\end{equation}
\noindent
where $x^{\rm \alpha}=(t,r,\theta,\phi)$ are the coordinates of a given photon, $\frac{dx^{\rm \alpha}}{d\lambda}=u^{\rm \alpha}$ its 4-velocity with $u\cdot u=0$, and $\lambda$ is an affine parameter. Here, the Christoffel symbols $\Gamma^{\rm \alpha}_{\rm \beta\gamma}$ have been expanded to first order in $\epsilon$ neglecting terms of order $\epsilon a$.

\section{PROPERTIES OF THE QUASI-KERR METRIC}

In this section, we analyze some of the properties of the quasi-Kerr metric. Glampedakis \& Babak (2006) obtained analytical expressions for equatorial orbits (where the Hamilton-Jacobi equation separates) as well as approximations for generic orbits in the vicinity of the black hole using canonical perturbation theory. Abramowicz et al. (2003) calculated approximate expressions for circular geodesics in the Hartle-Thorne metric (see, however, Berti et al. 2005 and Glampedakis \& Babak 2006).
We use, instead, a numerical integration of the geodesic equation for photons and particles in order to study the properties of orbits that determine observables in the electromagnetic spectrum.

All expressions in this section are calculated to the quadrupole order, i.e., they have been expanded in $\epsilon$ to first order, and terms of order $\epsilon a$ were neglected. Note that (following Glampedakis \& Babak 2006) we do not expand in the spin parameter $a$. This way we leave the Kerr part of all expressions unchanged and can study deviations from the Kerr metric by only changing the quadrupole moment.

\subsection{The Event Horizon and Static Limit}

We calculate the locations of the event horizon and of the static limit directly from the metric (\ref{qKerr}). We find the static limit by solving
\begin{equation}
g_{\rm tt}=0
\end{equation}
for $(r,\theta)$. The corresponding condition for the horizon is
\begin{equation}
g_{\rm t\phi}^2-g_{\rm tt}g_{\rm \phi\phi}=0.
\label{horizon}
\end{equation}

\begin{figure}[t*]
\begin{center}
\psfig{figure=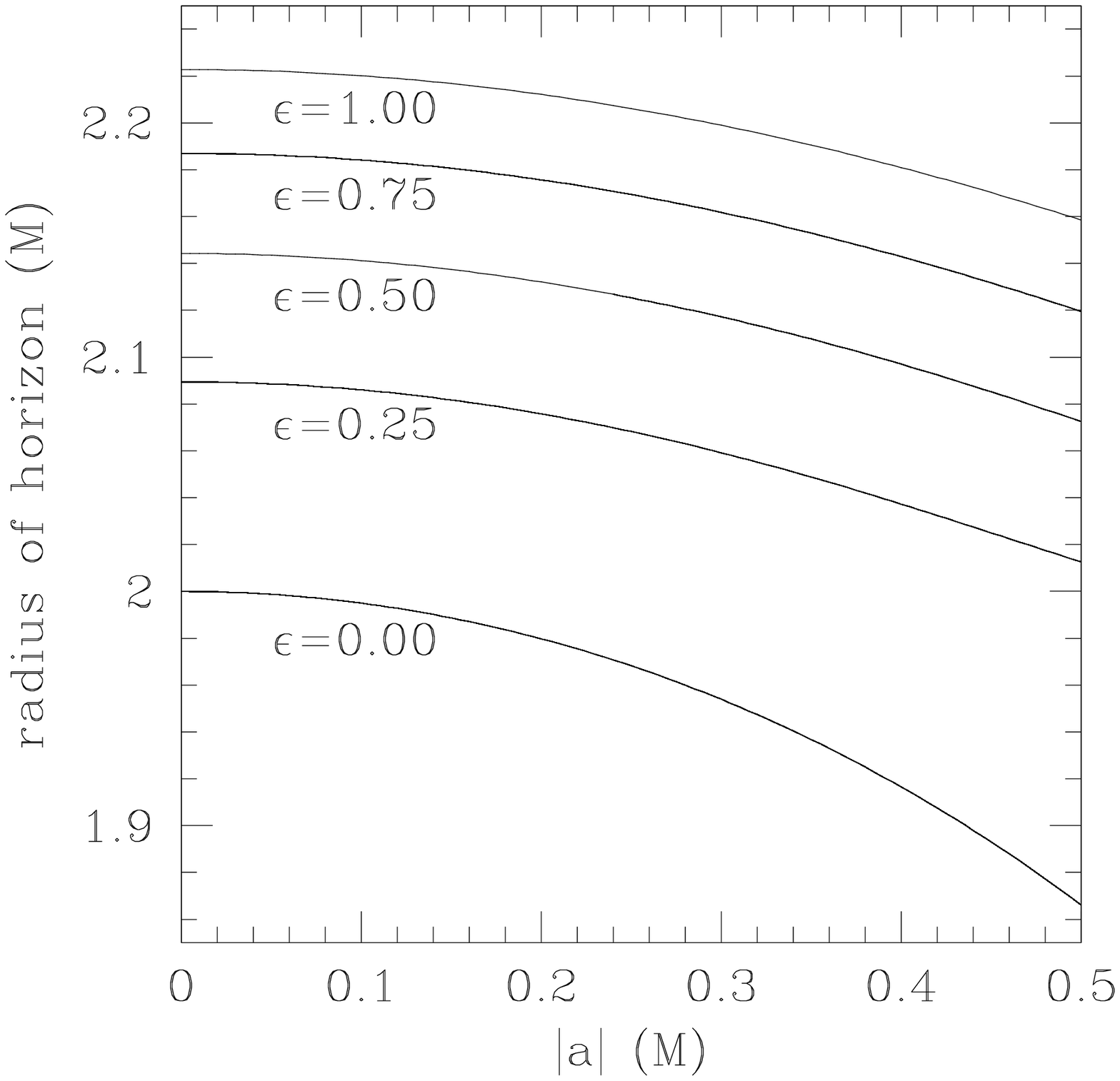,height=3.in}
\psfig{figure=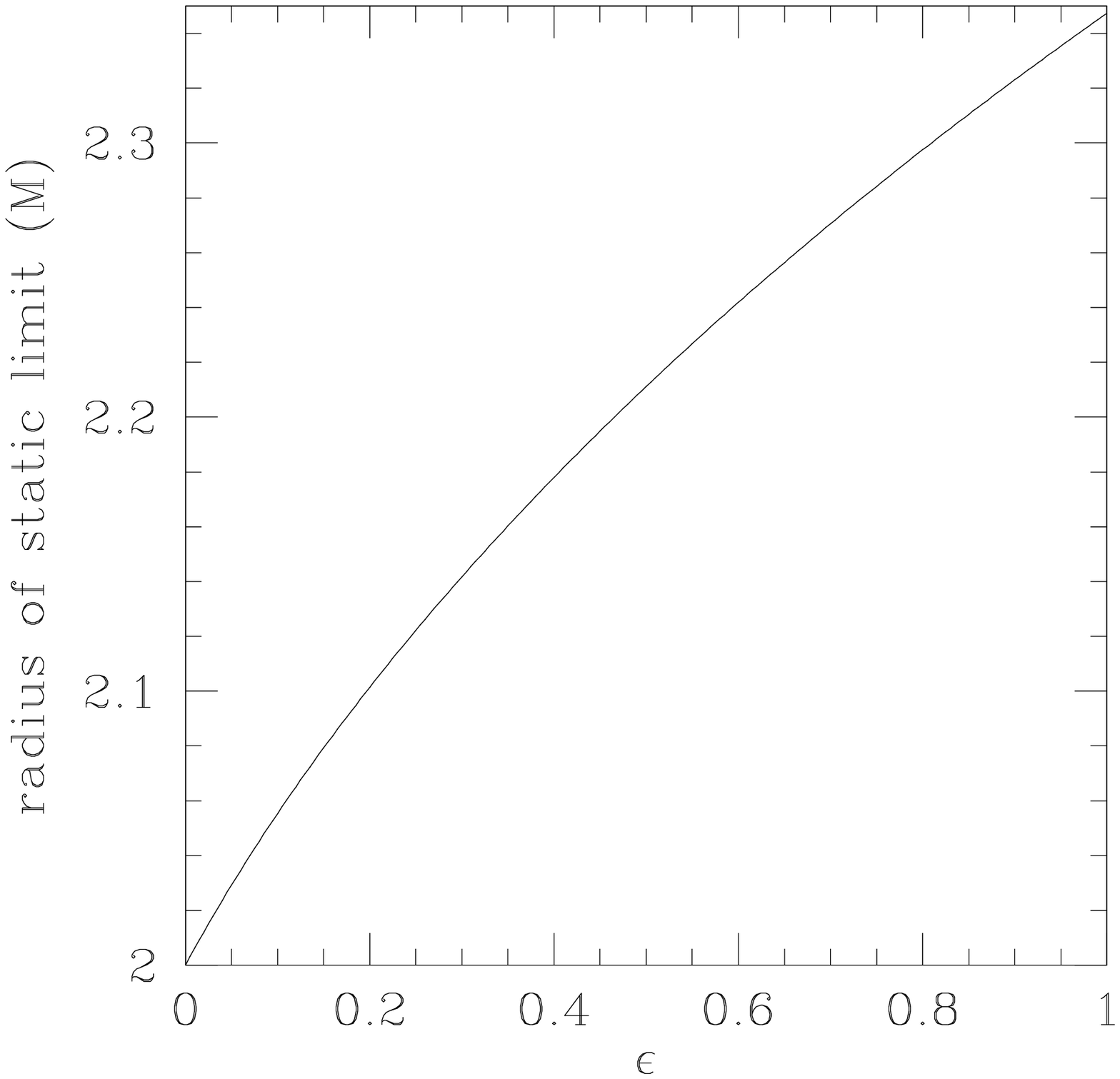,height=3.in}
\end{center}
\caption{{ \em (Left:)} The location of the event horizon in the equatorial plane of a quasi-Kerr black hole as a function of the absolute value of the spin $a$ for several values of the quadrupolar correction parameter $\epsilon$. Increasing the value of the spin decreases the radius of the event horizon, while increasing the value of the parameter $\epsilon$ increases that radius. {\em (Right:)} The equatorial static limit of a quasi-Kerr black hole as a function of the quadrupolar correction parameter $\epsilon$. A larger value of $\epsilon$ increases the radius of the static limit. In the equatorial plane, the static limit is independent of the spin $a$.}
\label{horizonstaticlimit}
\end{figure}

In Figure~\ref{horizonstaticlimit} we plot the radius of the event horizon in the equatorial plane as a function of the spin $a$ for several values of the parameter $\epsilon$. We also plot the radius of the static limit in the equatorial plane as a function of the parameter $\epsilon$. The horizon generally lies outside of the perturbative domain of the quasi-Kerr metric. If higher order terms are included, the radius of the event horizon might be slightly shifted or may even not exist at all (c.f., Manko \& Novikov 1992). The radius of the horizon increases with increasing positive values of the parameter $\epsilon$. If $\epsilon$ is negative, equation (\ref{horizon}) has no solution in general indicating the existence of a naked singularity. The radius of the horizon decreases as the spin $a$ gets larger as expected from the Kerr part of the metric. The static limit in the equatorial plane is independent of the spin $a$ and increases with increasing values of the parameter $\epsilon$. It likewise lies in the nonperturbative domain of the metric, but it gives a first insight into the importance of the quadrupolar correction parameter $\epsilon$ for relativistic boosting (c.f. Section \ref{releffects}). Note that the horizon is not regular for values of the angle $\theta\sim\pi/4$ or $\theta\sim3\pi/4$.

\subsection{The Innermost Stable Circular and Marginally Bound Orbits}

We now derive analytic expressions for the energy and angular momentum of massive particles on a circular equatorial orbit around a black hole in the quasi-Kerr metric. From these, we calculate the respective radii of the marginally bound orbit and of the innermost stable circular orbit (ISCO). The derivations are similar to the ones in Bardeen (1973) for the Kerr metric. All expressions are expanded to first order in $\epsilon$ omitting terms of order $\epsilon a$.

The quasi-Kerr metric is stationary and axisymmetric, and therefore generally admits three conserved quantities. For a particle with four-momentum
\begin{equation}
p^{\rm \alpha}=\mu\frac{dx^{\rm \alpha}}{d\tau}
\end{equation}
these are its rest mass $\mu$, energy $E=-p_{\rm t}$, and angular momentum about the $z$-axis $L_{\rm z}=p_{\rm \phi}$. Solving the equation
\begin{equation}
p\cdot p=-\mu^2
\end{equation}
in the equatorial plane for the radial derivative and substituting the constants of the motion yields
\[
r^3\left( \frac{dr}{d\tau} \right)^2 \equiv  R(r)
\]
\[
\equiv -4aMEL_{\rm z}-(r-2M)L_{\rm z}^2-\mu^2r[a^2+r(r-2M)]+[r^3+a^2(r+2M)]E^2 
\]
\[
- \frac{5\epsilon}{16M^2r} \bigg\{ 2M[\mu^2r^2(2M^3+4M^2r-9Mr^2+3r^3) 
\]
\[
+ 2(3M^3-2M^2r-6Mr^2+3r^3)L_{\rm z}^2]
\]
\begin{equation}
\left. +3r(r-2M)[-\mu^2r^3(r-2M)+2(M^2+Mr-r^2)L_{\rm z}^2]\ln\left(\frac{r}{r-2M} \right)    \right\}.
\end{equation}
We obtain the expressions for $E$ and $L_{\rm z}$ by solving the system of equations
\begin{equation}
R(r)=0,
\end{equation}

\begin{equation}
\frac{d}{dr}R(r)=0.
\end{equation}
After a nontrivial amount of algebra we find the energy
\[
\frac{E}{\mu}=\frac{ r^{3/2}-2Mr^{1/2}\pm aM^{1/2} }{ r^{3/4}\sqrt{r^{3/2}-3Mr^{1/2}\pm2aM^{1/2}} }
\]
\[
- \frac{ 5\epsilon }{ 32M^2r^{3/2}(r-3M)^{3/2} } \bigg[ 2M(6M^4+14M^3r-41M^2r^2+27Mr^3-6r^4)
\]
\begin{equation}
\left. +r^2(6r^3-33Mr^2+66M^2r-48M^3)\ln\left( \frac{r}{r-2M} \right) \right]
\label{energy}
\end{equation}
and angular momentum about the $z$-axis
\[
\frac{L_{\rm z}}{\mu}=\pm\frac{ M^{1/2}(r^2\mp2aM^{1/2}r^{1/2}+a^2) }{ r^{3/4}\sqrt{r^{3/2}-3Mr^{1/2}\pm2aM^{1/2}} }
\]
\[
\mp \frac{ 5\epsilon }{ 32M^{5/2}(r-3M)^{3/2} } \bigg[ 2M(6M^4-7M^3r-16M^2r^2+12Mr^3-3r^4) 
\]
\begin{equation}
\left. +3r(r^4-5Mr^3+9M^2r^2-2M^3r-6M^4)\ln\left( \frac{r}{r-2M} \right) \right].
\label{lz}
\end{equation}
For $\epsilon=0$ these expressions are identical to equations (17) and (18) in Bardeen (1973).

From the equation (\ref{energy}) for the energy we obtain the radius of the marginally bound orbit by numerically solving
\begin{equation}
\frac{E}{\mu}=1
\label{Eeqrmb}
\end{equation}
and of the innermost stable circular orbit (ISCO) from
\begin{equation}
\frac{dE}{dr}=0.
\label{Eeqisco}
\end{equation}
Similar calculations have also been performed by Shibata \& Sasaki (1998) and Berti \& Stergioulas (2004).

\begin{figure}[t]
\begin{center}
\psfig{figure=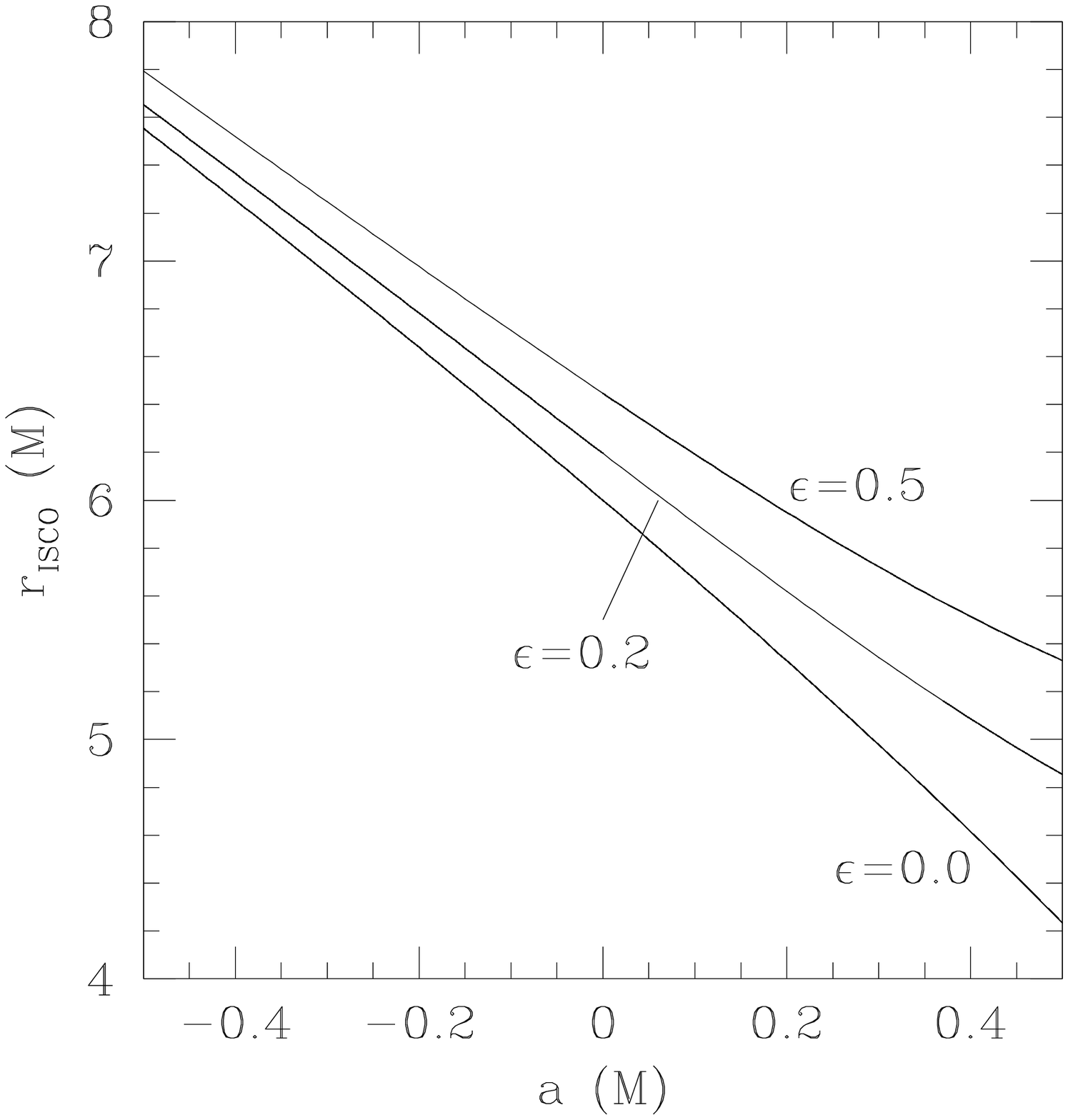,height=3.0in}
\psfig{figure=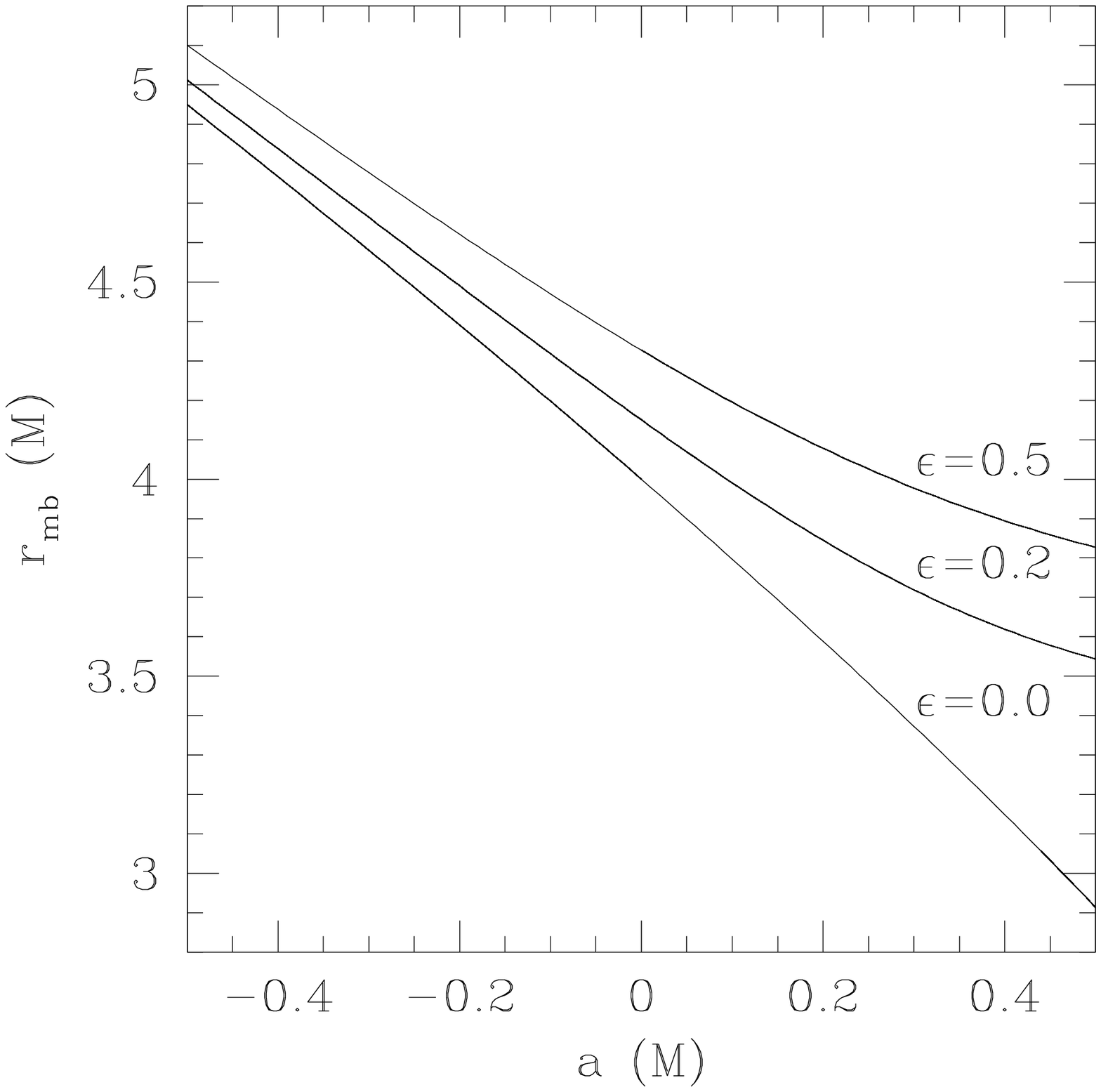,height=3.0in}
\end{center}
\caption{The radius of {\em (left)} the innermost stable circular orbit and {\em (right)} the marginally bound orbit for a massive particle orbiting a quasi-Kerr black hole as a function of the spin $a$ for several values of the quadrupolar correction parameter $\epsilon$. In both cases, a larger spin causes the respective orbits to be closer to the black hole, while an increasing value of the parameter $\epsilon$ moves the respective orbits to larger radii.}
\label{iscormb}
\end{figure}

Figure~\ref{iscormb} shows the radius of the innermost stable circular orbit, $r_{\rm ISCO}$, and the radius of the marginally bound orbit, $r_{\rm mb}$, for massive particles as a function of the spin $a$ for several values of the parameter $\epsilon$. Both radii increase with increasing positive $\epsilon$ and decreasing spin $a$. If $\epsilon$ is significantly smaller than zero, the expansions of the energy and the angular momentum in expressions (\ref{energy}) and (\ref{lz}) are not well defined, because the correction proportional to $\epsilon$ diverges at $r=3M$. Therefore, we only plot the radius of the ISCO and of the marginally bound orbit for values of the parameter $\epsilon\geq0$.

The ISCO is of special importance, because it determines the inner edge of the accretion disk of the black hole. If a massive particle from the accretion disk enters the region where $r<r_{\rm ISCO}$, it will quickly plunge into the black hole. In the Kerr geometry, the ISCO only depends on the mass and the spin of the black hole. A measurement of this radius for a black hole with known mass therefore allows for the spin to be measured (e.g., Brenneman \& Reynolds 2006; Zhang et al. 1997; Shaffee et al. 2006). In the quasi-Kerr metric, it also depends on the quadrupole moment (\ref{qradmoment}) and therefore on $\epsilon$. The significant dependence of the ISCO on the quadrupolar parameter $\epsilon$ will have a strong impact on many astrophysical applications such as black-hole images, iron lines, or disk spectra (see below).

\subsection{Photon Trajectories}

\begin{figure*}[t]
\begin{center}
\psfig{figure=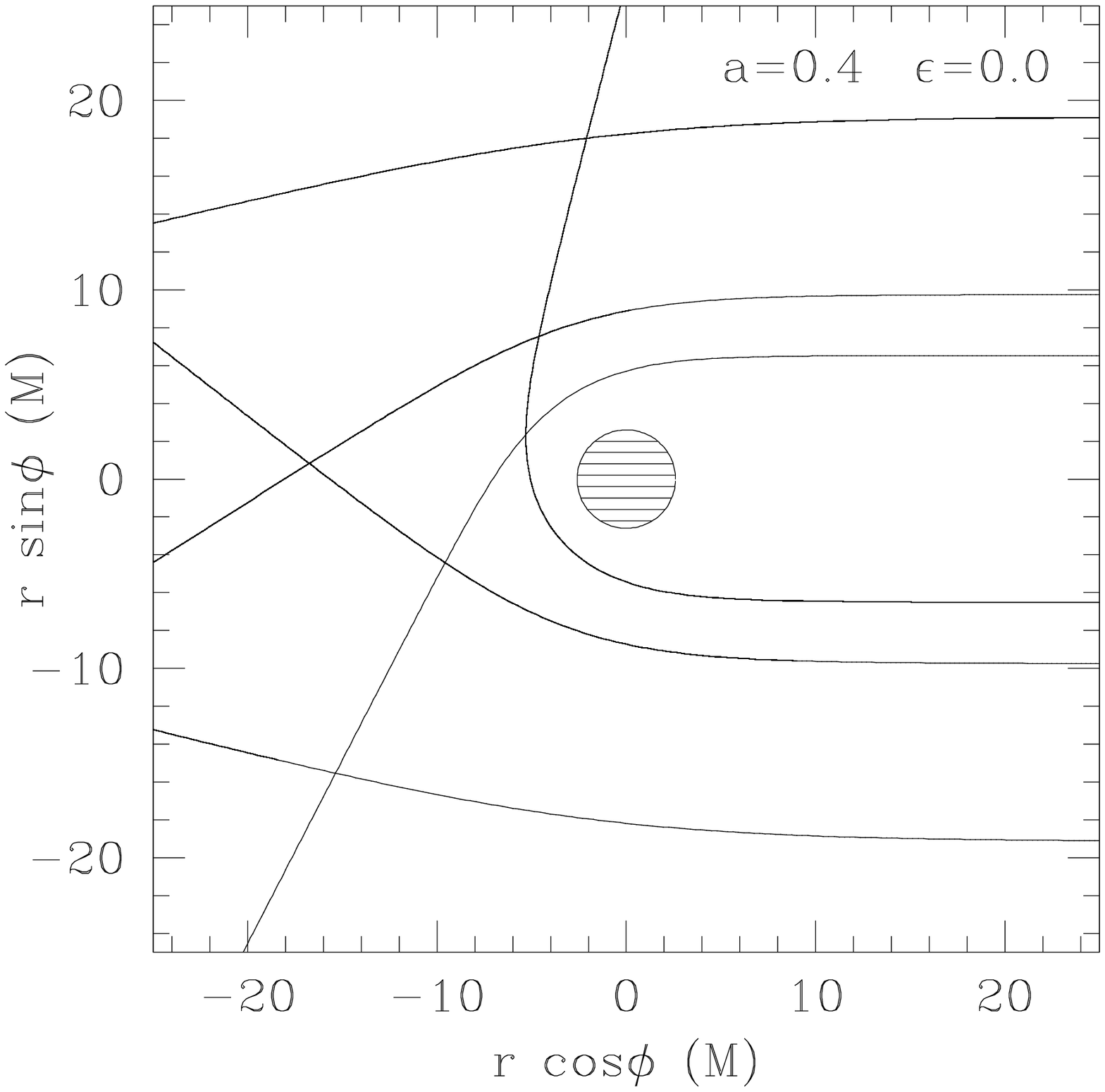,height=2.in}
\psfig{figure=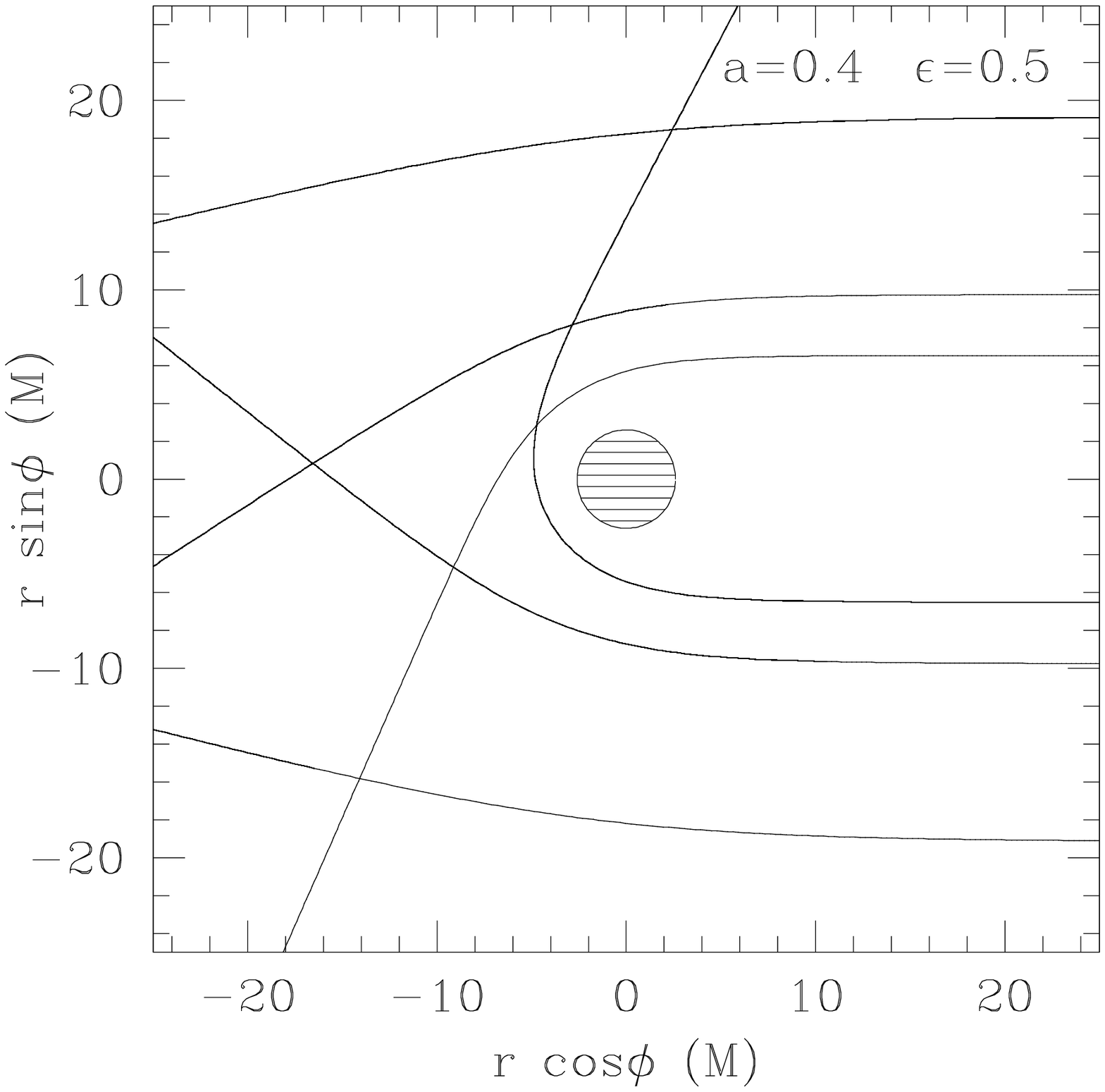,height=2.in}
\psfig{figure=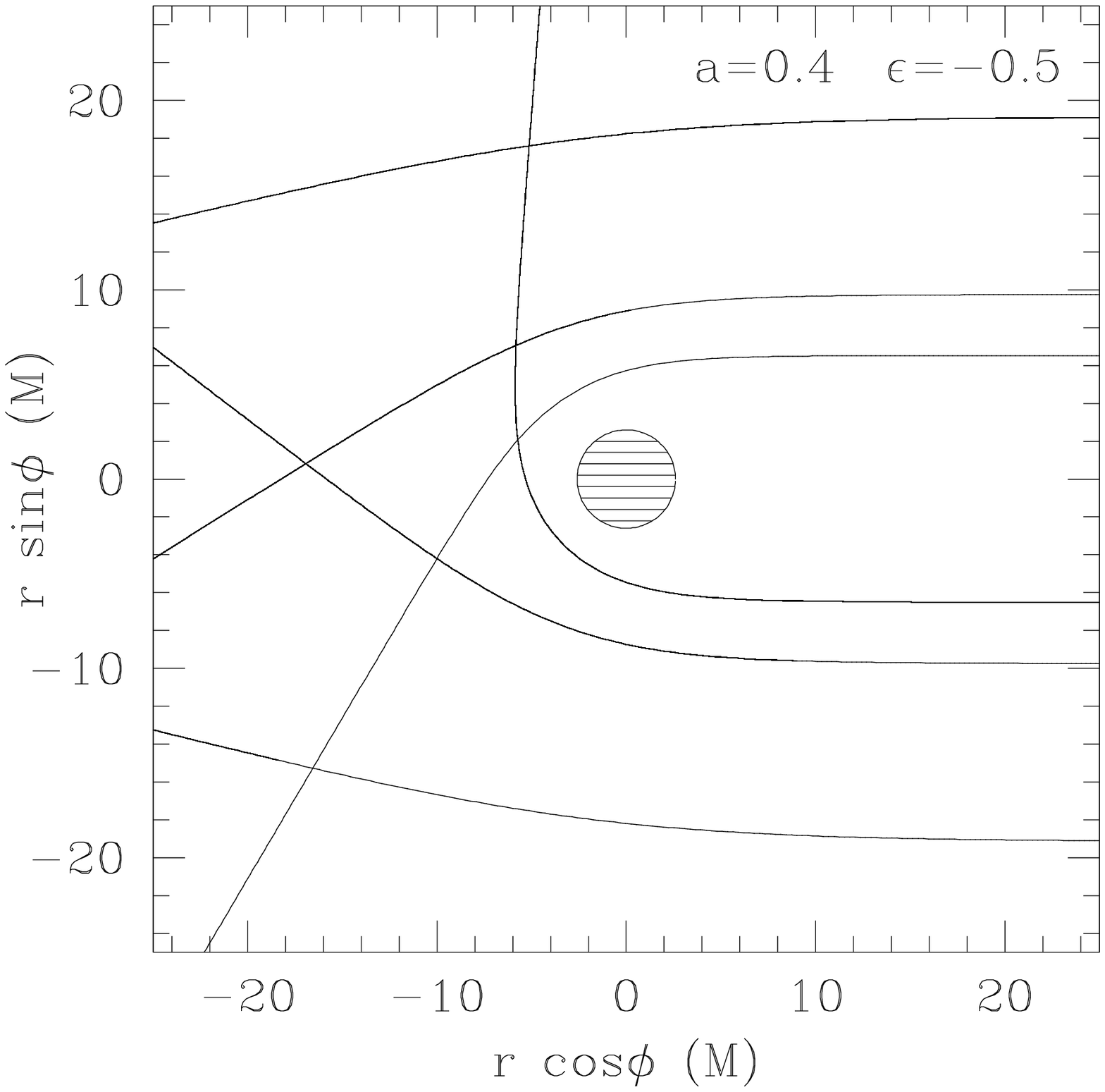,height=2.in}
\end{center}
\caption{Several equatorial photon trajectories around a Kerr black hole {\em (left)} and quasi-Kerr black holes with quadrupolar parameter $\epsilon=0.5$ {\em (center)} and $\epsilon=-0.5$ {\em (right)}, respectively. The spin parameter has a value of $a=0.4$ in all cases. Photons approach the black hole from the right. The trajectories experience the strongest bent the closer they approach the black hole. Frame dragging shifts orbits in the direction of (the counter-clockwise) rotation of the black hole, while an enhanced/decreased quadrupole moment increases/decreases the light bending in the immediate vicinity of the black hole. The shaded region marks the excluded domain $r<2.6M$.}
\label{skorbits}
\end{figure*}

\begin{figure}[h]
\centerline{
\includegraphics[width=0.5\textwidth]{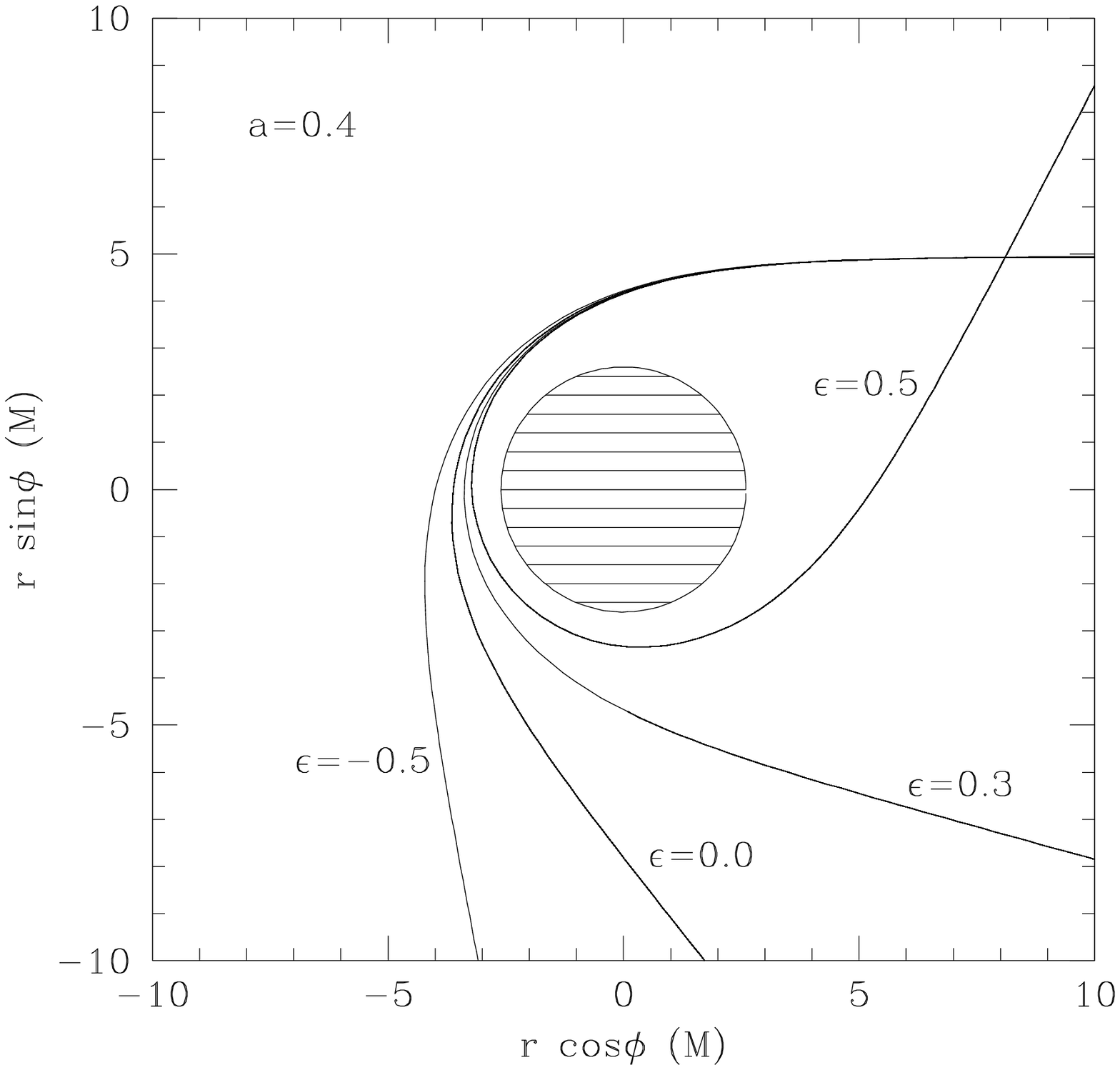}
}
\caption{Scattering of equatorial photon trajectories by a quasi-Kerr black hole with a (counter-clockwise) spin $a=0.4$ for several values of the quadrupolar correction parameter $\epsilon$. The photons approach the black hole from the right. The shaded region marks the excluded domain $r<2.6M$. The larger the value of the parameter $\epsilon$, the stronger the amount of light bending.}
\label{scatterorbits}
\end{figure}

Close to the black hole, the quadrupole moment becomes important and affects the trajectories of photons and massive particles in the accretion disk. Since the quasi-Kerr metric is of Petrov-type D in the equatorial plane, these orbits remain equatorial. We integrated the geodesic equation (\ref{geodesiceq}) numerically using a Runge-Kutta method with adaptive stepsize for several equatorial photon orbits.

Figure~\ref{skorbits} illustrates the impact of the interplay between the black hole's spin and quadrupole moment on the equatorial orbits. A positive spin $a$ corresponds to counter-clockwise rotation of the black hole. A collection of light rays approaches the black hole parallel to the $x$-axis from the right. The figure is scale invariant with respect to the black-hole mass $M$, and all quantities are expressed in appropriate powers of the mass $M$. The shaded region corresponds to the excluded domain $r<2.6M$. 

For the Kerr black hole, trajectories are shifted due to the usual frame dragging. For the quasi-Kerr black holes, a change of the quadrupole moment away from its value in the Kerr metric either increases or decreases the amount of light bending. For a positive value of the quadrupolar correction $\epsilon$, trajectories are shifted further in the clockwise direction. For a negative value of the parameter $\epsilon$, the reverse is the case for trajectories that approach the black hole closely. As expected, in both cases, an alteration of the quadrupole moment only effects trajectories in the immediate vicinity of the black hole, while photon trajectories further away from the black hole are dominated by the spin.

In Figure \ref{scatterorbits} we show the effect of varying the quadrupole moment on one trajectory in particular. We held the spin fixed at $a=0.4$ and show photon trajectories for $\epsilon=-0.5,~0.0,~0.3,~0.5$. The effect of the quadrupole moment becomes very apparent once the photons approach a radius of $r\sim3M$. A positive value of $\epsilon$ increases the amount of light bending, while that effect deceases for negative values of $\epsilon$ compared to the Kerr trajectory. Note that for this particular orbit, the asymptotic azimuthal angle $\phi_{\rm \infty}$ differs by more than $\pi/2$ for the orbits with $\epsilon=-0.5$ and $\epsilon=0.5$, respectively. These effects will significantly alter the shadows of black holes as well as the details of accretion flow spectra.

\subsection{Gravitational Redshift and Lorentz Boosts}

\label{releffects}

\begin{figure}[h]
\centerline{
\includegraphics[width=0.5\textwidth]{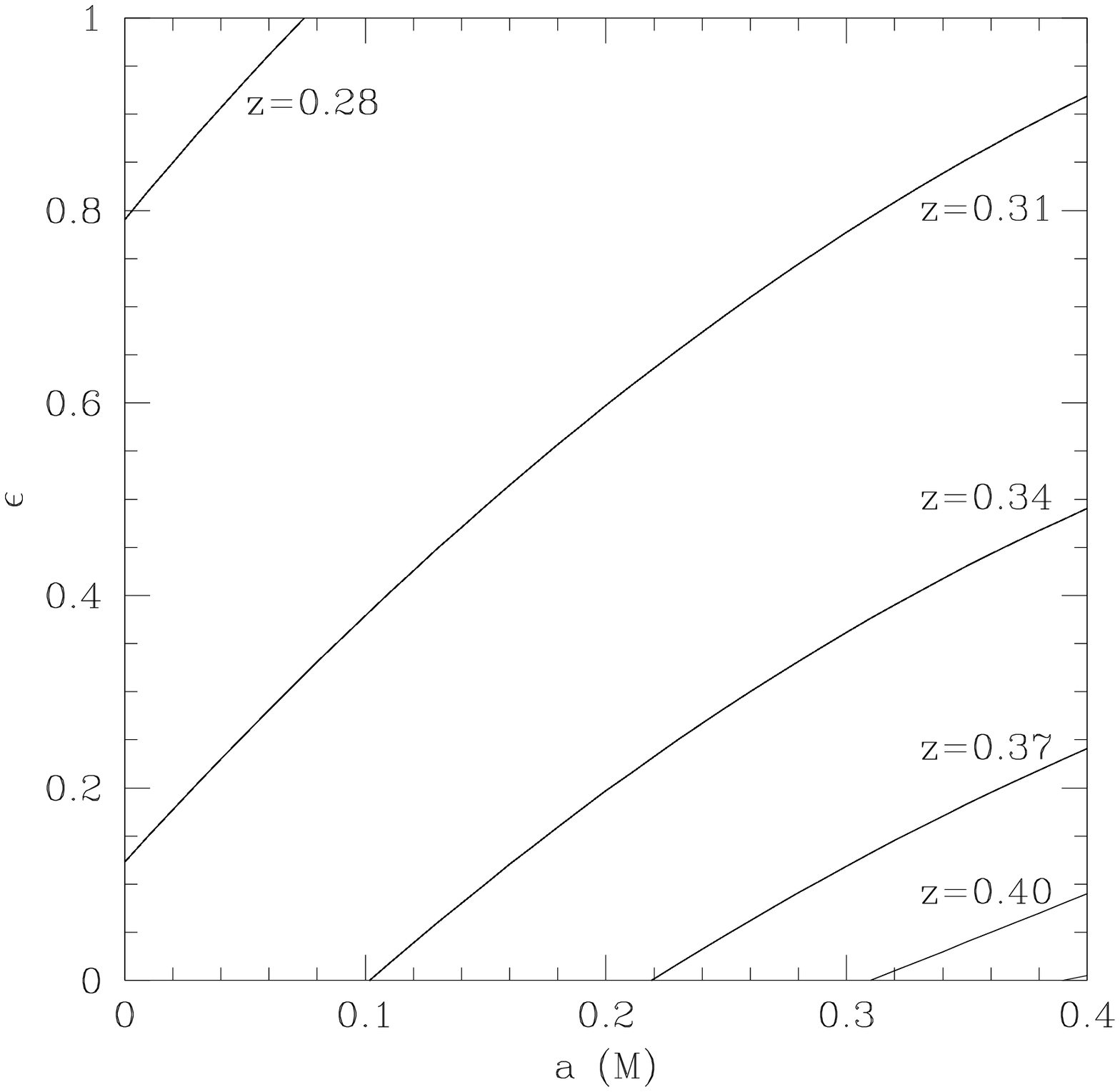}
}
\caption{Contours of constant redshift $z$ experienced by a photon emitted by a particle on the (prograde) ISCO as viewed by an observer at rest at infinity as a function of black-hole spin $a$ and the quadrupolar correction parameter $\epsilon$. In all cases, the impact parameter of the photon is set to $b=1$. The higher the spin, the higher the redshift. Increasing the values of the parameter $\epsilon$ decreases the redshift.}
\label{redshift}
\end{figure}

In this section we investigate the dependence of gravitational redshift and Lorentz boosting of photons emitted in the quasi-Kerr spacetime on the quadrupolar parameter $\epsilon$. These effects are critical for predicting the observational appearance of astrophysical processes in the vicinity of black holes. In particular, they determine the flux ratios between the approaching and receding parts of accretion flow images, as well as the width of relativistically broadened iron lines.

We calculate the redshift at infinity of a photon that is emitted by a particle on the ISCO. The coordinate velocity of a massive particle on a circular orbit in the equatorial plane with radius $r$ is given by (Glampedakis \& Babak 2006)
\[
u^{\rm t}=\frac{1}{\Delta} \left[ E(r^2+a^2)+\frac{2Ma}{r}(aE-L_{\rm z}) \right] - \epsilon \left( 1-\frac{2M}{r} \right)^{-1}f_3(r)E
\]
\begin{equation}
u^{\rm \phi}=\frac{1}{\Delta} \left[ \frac{2M}{r}(aE-L_{\rm z})+L_{\rm z} \right] - \epsilon\frac{h_3(r)}{r^2}L_{\rm z},
\label{uisco}
\end{equation}
where
\[
f_3(r)=-\frac{5(r-M)}{8Mr(r-2M)}(2M^2+6Mr-3r^2)-\frac{15r(r-2M)}{16M^2}\ln \left( \frac{r}{r-2M} \right)
\]
\begin{equation}
h_3(r)=\frac{5}{8Mr}(2M^2-3Mr-3r^2)+\frac{15}{16M^2}(r^2-2M^2)\ln \left( \frac {r}{r-2M} \right).
\end{equation}
The orbital energy $E$ and angular momentum about the $z$-axis $L_{\rm z}$ are given by equations (\ref{energy}) and (\ref{lz}) and have been normalized to the particle's rest mass $\mu$. In these expressions, the radius is evaluated at $r=r_{\rm ISCO}$.

The redshift of a photon at infinity is defined by (e.g., Cunningham 1975)
\begin{equation}
\frac{1}{1+z}=\frac{p_{\rm t}}{p_{\rm e}\cdot u},
\end{equation}
which includes both the gravitational redshift and the Doppler shift. Here, $p_{\rm t}$ is the photon energy with respect to an observer at rest at infinity, $p_{\rm e}$ the 4-momentum of the photon at its emission, and $u$ is the 4-velocity of the emitting particle given by equation (\ref{uisco}). Therefore, the redshift is equal to
\begin{equation}
z=u^{\rm t}-b u^{\rm \phi}-1,
\label{totalz}
\end{equation}
where
\begin{equation}
b=-\frac{ p_{\rm \phi} }{ p_{\rm t} }
\end{equation}
is the impact parameter of the emitted photon.

Figure \ref{redshift} shows contours of constant redshift $z$ at infinity of a photon with a nominal value of the impact parameter $b=1$ emitted by a particle on the (direct) ISCO around the black hole. The respective values of the redshift range from $z=0.28$ for the line in the top left corner to $z=0.40$ along the line in the bottom right corner in steps of $0.03$. The higher the spin of the black hole, the larger the redshift of the photon, because the radius $r_{\rm ISCO}$ decreases so that it costs the photon more energy to escape the gravitational attraction of the black hole. For increasing quadrupolar correction $\epsilon$ at a given spin $a$, however, the redshift decreases, because the radius $r_{\rm ISCO}$ (c.f. Figure \ref{iscormb}) and the azimuthal velocity $u_{\rm \phi}$ (as well as $u_{\rm t}$) increase.

\subsection{The Circular Photon Orbit}

In principle, we can obtain the radius of the circular photon orbit by inverting expressions (\ref{energy}) and (\ref{lz}). Once properly expanded to the quadrupole order, the photon radius occurs where the resulting expressions vanish. Unfortunately, in our case, this procedure leads to slightly different results when either the expression for the energy (\ref{energy}) or the one for the axial angular momentum (\ref{lz}) are inverted. This is a consequence of the expansion to first order in the parameter $\epsilon$ and higher order corrections are required. Instead, we obtain a numerical solution for the circular photon orbit in the following way.

\begin{figure}[h]
\centerline{
\includegraphics[width=0.5\textwidth]{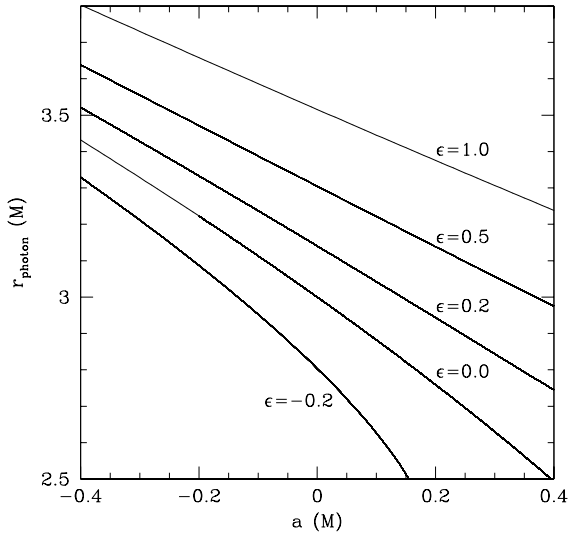}
}
\caption{The radius of the circular photon orbit $r_{\rm photon}$ as a function of the spin $a$ for several values of the quadrupolar correction parameter $\epsilon$. The radius $r_{\rm photon}$ decreases for increasing values of the spin due to frame dragging, while $r_{\rm photon}$ increases for increasing values of the parameter $\epsilon$.}
\label{photonorb}
\end{figure}

In order to calculate the radius of the circular photon orbit numerically, we solved the geodesic equation (\ref{geodesiceq}) in the equatorial plane for a photon with a purely azimuthal initial 3-velocity for various initial radii and determined the initial radius for which the photon trajectory changed from being unbound to being bound. In Figure~\ref{photonorb} we plot the radius of the circular photon orbit $r_{\rm photon}$ as a function of the spin $a$ for several values of the parameter $\epsilon$. The radius $r_{\rm photon}$ decreases for increasing values of the spin due to frame dragging, while $r_{\rm photon}$ increases for increasing values of the parameter $\epsilon$.

\section{CONCLUSIONS}

In this series of papers, we investigate a framework for testing the no-hair theorem with observations of black holes in the electromagnetic spectrum. Our framework can be viewed as either a null-hypothesis test of the no-hair theorem within general relativity or as a self-consistent test of general relativity itself in the strong-field regime. If the multipole moments of a black-hole candidate are measured to be different from the moments of the Kerr metric, there are two possibilities. If general relativity is assumed, the astrophysical object is not a black hole. But if it is known to possess a horizon, then both the no-hair theorem and general relativity are incorrect. If the measured multipole moments coincide with the Kerr multipole moments, general relativity and the no-hair theorem may be correct or not. A definite answer from the extraction of the multipole moments alone is not possible, because other theories of gravity likewise predict the Kerr metric as a black-hole spacetime (Psaltis et al. 2008).

In this first paper, we formulated our tests based on the quasi-Kerr metric (Glampedakis \& Babak 2006) which contains an independent quadrupole moment and deviates smoothly from the Kerr metric at the quadrupole order. Since the no-hair theorem admits exactly two independent multipole moments, a measurement of three moments allows us to test the no-hair theorem (Ryan 1995). General-relativistic black holes must have the multipole structure of the Kerr metric. If a different set of multipoles is detected, then the compact object cannot be a black hole within general relativity.

We estimated the range of validity of the quasi-Kerr metric and demonstrated the dependence of various properties of this spacetime on both the spin and the quadrupole moment of the black hole. We analyzed in detail the effects of light bending, photon redshift, and the respective locations of the ISCO and the circular photon orbit, all of which are of critical importance for observables of accretion flows around astrophysical black holes. In particular, we showed that the radius of the ISCO and the amount of gravitational lensing experienced by photons are altered significantly for already moderate changes of the quadrupole moment.

We identified several observational approaches within our framework that will test the no-hair theorem within the next few years. Among these are imaging observations of accretion flows around black holes and, especially, Sgr A* using VLBI techniques, as well as the precise measurements of the spectra of iron lines and accretion disks from AGN with IXO. In the following papers we will explore in detail the prospects of testing the no-hair theorem with such observations.

We thank Avery Broderick, Abraham Loeb, Scott Hughes, Kostas Glampedakis, Emanuele Berti, Nico Yunes, Sarah Vigeland, and Daniel Marrone for helpful discussions. We also thank Sarah Vigeland and Scott Hughes for sharing their paper with us in advance of publication. DP thanks the ITC at the Harvard-Smithsonian Center for Astrophysics for their hospitality. This work was supported by the NSF CAREER award NSF 0746549.

\end{document}